# Ultrafast control of braiding topology in non-Hermitian metasurfaces

Yuze Hu[1,2,†], Mingyu Tong[1,3,4,5,†], Ziheng Ren[6], Fujia Chen[1,3,4], Qiaolu Chen[1,3,4,5], Hongsheng Chen[1,3,4,5,*], Tian Jiang[2,*], Yihao Yang[1,3,4,*]

[1] Interdisciplinary Center for Quantum Information, State Key Laboratory of Extreme Photonics and Instrumentation, ZJU-Hangzhou Global Scientific and Technological Innovation Center, Zhejiang University, Hangzhou 310027, China.
[2] Institute for Quantum Science and Technology, College of Science, National University of Defense Technology, Changsha, 410073, China
[3] International Joint Innovation Center, The Electromagnetics Academy at Zhejiang University, Zhejiang University, Haining 314400, China.
[4] Key Lab. of Advanced Micro/Nano Electronic Devices & Smart Systems of Zhejiang, Jinhua Institute of Zhejiang University, Zhejiang University, Jinhua 321099, China.
[5] Shaoxing Institute of Zhejiang University, Zhejiang University, Shaoxing 312000, China.
[6] College of Advanced Interdisciplinary Studies, National University of Defense Technology, Changsha, 410073, China.
[†] These authors contributed equally to this work.
*Corresponding author. Email: yangyihao@zju.edu.cn (Y.Y.); tjiang@nudt.edu.cn (T.J.); hansomchen@zju.edu.cn (H.C.)

## Abstract

The mathematical theory of braids, influential across scientific disciplines, has emerged as a compelling strategy for light manipulation. Existing approaches to creating braids in photonics, whether in momentum-space bandstructures or real-space fields, often face limitations associated with static nature of devices and lack of tunability. Here, we experimentally demonstrate ultrafast control of eigen-spectrum braids of Jones matrices within mere picoseconds, in reconfigurable non-Hermitian metasurfaces. The Jones matrices of the metasurface exhibit a complex eigen-spectrum that braids in the three-dimensional eigenvalue-frequency space, thereby creating arbitrary elements within the two-string braid group, $\mathbb{B}_2$. By exciting the photoconductive semiconductor terahertz metasurface with a femtosecond infrared pulse, we achieve ultrafast switching of the braids, transitioning from the Solomon link to either the Trefoil knot or Hopf link. Our approach serves as a pivotal tool for elucidating non-trivial topology of braids and studying ultrafast topological optoelectronics.

# Main Text

Braids, as fundamental geometrical configurations, manifest ubiquitously in natural phenomena from fluid dynamics to biological DNA structures [1,2]. For enhance visualization, braids can be seamlessly transformed into a solid torus ($T^2$) by uniting their boundaries, thereby transforming them into closed knot components. These components are distinguished by their unique nonzero integer knot invariants, indicative of distinct braiding topologies [3]. The rich topologies have spanned a broad spectrum of physics ranging from real space to momentum space, which encompass various quantum effects in high-energy [4] and condensed matter physics [5,6], as well as exotic phenomena in classical fields such as mechanics [7], acoustics [8], and photonics [9-13].

In particular, the exploration of braids as structures for electromagnetic waves is currently under active investigation in photonics. In momentum space, the closed trajectories formed by the degeneracies of energy bands in Hermitian systems, or the complex energy bands in non-Hermitian systems, give rise to a variety of knot configurations [14-17]. In real space, the complicated light field can be sculpted into vortex knots and linked shapes, using spatial light modulators or intricately engineered nanostructures [18-22], which have also been underscored the practicality as a means to encode information [10,23]. In parallel, ultrafast switching photonic states hold immense potential for revolutionizing photonic technologies, including routing, modulation, and frequency conversion, crucial for advancements in information processing, telecommunications, imaging, and sensing [24-28]. Nevertheless, it is not obvious how one can flexibly transform braiding topology without statically replacing devices, leaving the significant potential for exploring and controlling a large number of braiding topology effects in ultrafast photonics.

Here, we theoretically propose and experimentally demonstrate a framework of ultrafast switching eigen-spectrum braids of the transmitted Jones matrices in reconfigurable non-Hermitian metasurfaces, achieved within a few picoseconds. Our metasurface employs four coupled semiconductor-hybrid terahertz plasmonic split ring resonators (SRRs), where tunability is achieved through transient nonradiative loss modulation [29-31]. The transmitted Jones matrices of the metasurface exhibit complex eigen-spectrum that braids in the three-dimensional (3D) eigenvalue-frequency space, which can form arbitrary elements in the two-string braid group, $\mathbb{B}_2$. By exciting the metasurface with a femtosecond infrared pulse, we experimentally evidence various ultrafast transitions among Solomon link, Trefoil knot, and Hopf link when the system traverses chiral exceptional points (EPs).

Our study begins with the analysis of transmitted Jones matrices in a four-resonators metasurface under external incidence, providing a simplified yet comprehensive representation for analytical examination (see Figs. 1a-b). The system's transmitted Jones matrix, expressed as $\hat{\mathbf{H}}_\mathrm{T}(\omega) = [t_{xx}, t_{xy}; t_{yx}, t_{yy}] = F(\hat{\mathbf{H}}_\mathbf{M})$, exhibits two complex eigenvalues observable in the far-fields (see detailed expression of $F(\hat{\mathbf{H}}_\mathbf{M})$ in Supplementary Note 1). Here, $\omega = 2\pi f$ is the angular frequency, $\hat{\mathbf{H}}_\mathbf{M}$ is a $4 \times 4$ Hamiltonian describing the coupling and loss of the resonators, and $t_{uv}$ is a complex transmission for $u$-polarized outgoing wave with $v$-polarized incidence, $(u,v) \subseteq \{x,y\}$. Note that $t_{xy}$ and $t_{yx}$ is equivalent when the reciprocity of a metasurface is preserved. Moreover, $\hat{\mathbf{H}}_\mathrm{T}(\omega)$ can be written in terms of Pauli matrices, such that $\hat{\mathbf{H}}_\mathrm{T}(\omega) = d_0 \mathbf{I}_0 + d_x \boldsymbol{\sigma}_x + d_y \boldsymbol{\sigma}_y + d_z \boldsymbol{\sigma}_z$, with $d_0 = \dfrac{t_{xx} + t_{yy}}{2}$, $d_x = t_{xy}$, $d_y = 0$, and $d_z = \dfrac{t_{xx} - t_{yy}}{2}$.

The transmitted Jones matrix $\hat{\mathbf{H}}_\mathrm{T}(\omega)$ can be also viewed as a Hamiltonian, with the eigenstates representing the eigenpolarization states that remain unchanged after passing through the metasurface and the corresponding eigenvalues being the complex-valued transmission of the eigenstates [32-34]. An important feature of the Hamiltonian $\hat{\mathbf{H}}_\mathrm{T}(\omega)$ is that it is non-Hermitian and has two complex-valued eigenvalues. As the frequency ($\omega = 2\pi f$) evolves, each eigenvalue ($\lambda$) can form a string of a braid in the (Re($\lambda$), Im($\lambda$), $\omega$)-space, which is a three-dimensional space. Moreover, at frequencies far removed from the resonances, the eigenvalues and the corresponding eigenstates of $\hat{\mathbf{H}}_\mathrm{T}(\omega)$ tend to be identical. Consequently, the two ends of each braid can be united, forming a closed knot configuration (see details in Supplementary Text). Therefore, the two eigenvalues of $\hat{\mathbf{H}}_\mathrm{T}(\omega)$ in the eigenvalue-frequency space can be naturally described by the knot theory.

In our case, the number of strings in the braid is two, which is the minimal string number required to generate knots or link configurations. The topological invariant associated with these two-string braids corresponds to the braid group $\mathbb{B}_2$, recognized as the most elementary braid group. This particular braid group is uniquely characterized without reliance on a reference eigenvalue, unlike the knot invariant formulated for single-string scenarios which necessitates such a value (e.g., the knot invariant defined in the point-gap topology for the Hatano-Nelson Hamiltonian [35,36], requiring a reference eigenenergy). Thus, our two-string approach bears similarity to the concept of two-band complex-energy braiding [14,15]. Nevertheless, it should be noted that the framework of our braids operates within a space where Jones matrices' eigenvalues and frequency constitute the basis, different from the previous cases with a space defined by eigenenergies and momenta.

To better visualize the braiding topology, we apply a gauge transformation to $\hat{\mathbf{H}}_{\mathrm{T}}(\omega)$, $\hat{\mathbf{H}}_{\mathbf{T}}' = \boldsymbol{\sigma}_y^{-1}(\hat{\mathbf{H}}_{\mathrm{T}} - d_0 \mathbf{I}_0)\boldsymbol{\sigma}_y$. The gauged Hamiltonian $\hat{\mathbf{H}}_{\mathbf{T}}'(\omega)$ anti-commutes with $\boldsymbol{\sigma}_y$, exhibiting chiral symmetry that enforces the two eigenvalues inversion symmetric. Note that the gauged Hamiltonian can be directly measured and the gauge transformation does not change the braiding topology. By properly choosing the parameters away from the EPs, the two strings can satisfy the separable condition [37]. The topology of two separated strings can be classified by the braid group $\mathbb{B}_2$, which is isomorphic to the group of integers $\mathbb{Z}$ [3,38]. The corresponding integer topological invariant can be calculated

$$\mathcal{W} = \int_{-\infty}^{+\infty} \frac{d\omega}{2\pi i} \frac{d}{d\omega} \ln[\lambda_{\mathbf{1,2}}(w)], \qquad (1)$$

where $\lambda_{\mathbf{1,2}}(\omega) = \pm \frac{\sqrt{(t_{xx} - t_{yy})^2 + 4t_{xy}^2}}{2}$ are the eigenvalues of $\hat{\mathbf{H}}_{\mathrm{T}}'(\omega)$.

We are now poised to construct a theoretical framework central to our exploration of braiding topology transition in non-Hermitian metasurfaces. By setting $\omega_{p,m} = \omega_0 + \Delta\omega$ and $\omega_{q,n} = \omega_0 - \Delta\omega$, $\lambda_{\mathbf{1,2}}(\omega)$ lives on a 3D parameter space spanned by $(\Delta\gamma_{pm}', \Delta\gamma_{qn}', \Delta\omega) \in \mathbb{R}^3$. Our analysis reveals the emergence of two intersecting exceptional surfaces (ESs), depicted as red and blue planes in Fig. 1c, dividing the space into four distinct subspaces (i.e., subspaces i, ii, iii, iv). The interaction of these two strings, alternating from unbraiding to braiding $\mathcal{W}$ times, culminates in the formation of various topology entities, such as a Solomon link ($\mathcal{W} = 4$), a Trefoil knot ($\mathcal{W} = 3$), and a Hopf link ($\mathcal{W} = 2$), each corresponding to a specific manner of joining the strings into a torus structure [39]. The complex eigenvalue spectra within these subspaces are illustrated in Figs. 1d-h, where two enveloping bands display three characteristic braids corresponding to a Solomon link, a Trefoil knot, and a Hopf link, respectively (see details in Supplementary Text). These braids, stemming from the four subspaces, represent topologically unique non-Hermitian spectrum structures, distinguishable by their inability to be continuously transformed into one another without intersecting the ES, where string collapse occurs at a specific frequency. This theoretical framework thus delineates distinct braiding topology in the 1D complex spectrum, grounded in homotopy classification.

Given the close association of EP singularities with braid phase transitions, the eigen spectra of $\hat{\mathbf{H}}_T(\omega)$ are scrutinized by adjusting parameters $\Delta\gamma'_{pm}$ and $\Delta\gamma'_{qn}$. Spectral trajectories, as illustrated in Figs. 1d-h, exhibit varied braiding behaviors, particularly when disentangling two strings at different locations. Traversing ESs, as marked in Fig. 1c, is necessary for the disentangle process, culminating in eigenvalue degeneracy as demonstrated in the first column of Figs. 1e-f (see detailed parameters in Supplementary Note 1). Besides, the projected eigenstates of Jones matrix on a Poincaré sphere are particularly demonstrated, all showing closed loops in the whole spectral range. These closed loops of eigenstates are closely related to the compact space along frequency dimension defined for separable eigen spectra of Jones matrix. At singularities, it is found two eigenstates coalesce into the sphere's poles for either left- or right-chiral polarization, representing opposite chiral nature of two exceptional surfaces in Fig. 1c. Interestingly, the trajectory of the eigenstates on the Poincaré sphere also show different topology, i.e., a single loop (e.g., subspaces ii and iii) or double loops (e.g., subspaces i and iv), corresponding to links (Solomon and Hopf) and knot (Trefoil), respectively.

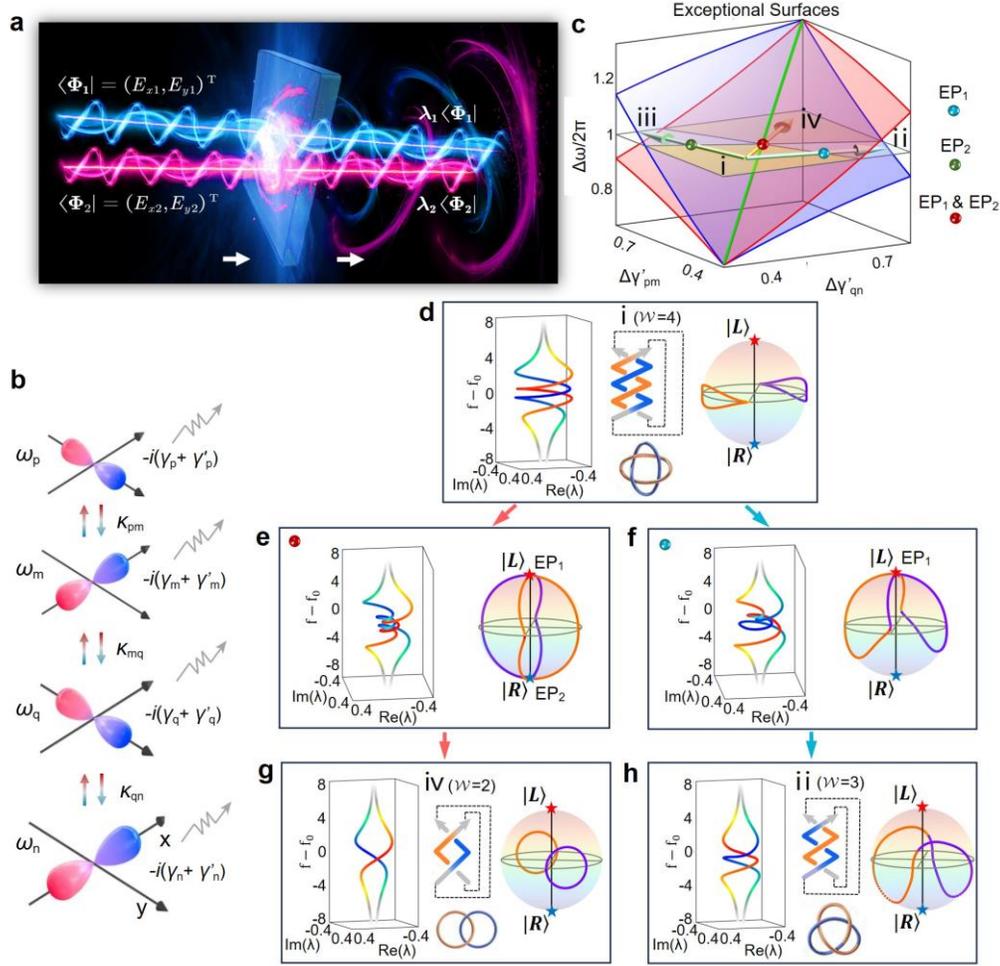

**Figure 1 | Eigenspectral braids of Jones matrix in non-Hermitian metasurfaces. a,** Schematic representation of dual-strings along the frequency dimension in the transmitted Jones matrix eigenspace. Each Jones matrix shows two eigenstates with corresponding complex eigenvalues. **b,** Resonator configuration featuring four dipole resonances coupled with two orthogonally polarized incident waves. The *y*-polarized resonators are marked as 'p', 'q', while the *x*-polarized ones are 'm' and 'n'. $\gamma$ and $\gamma'$ representing the radiative and non-radiative losses, respectively. **c,** Comprehensive phase diagram, showcasing various braid types and their transitions across two exceptional surfaces. It is delineated into four regimes, including Solomon knot (subspace i), Trefoil link (subspaces ii-iii), and Hopf link (subspace iv). The phase transitions are controllable by tuning the loss imbalances of $\Delta\gamma'_{qn}$ and $\Delta\gamma'_{pm}$. The green line represents the second order exceptional line. **d-h,** Frequency-resolved complex eigenvalues (left panel) and projected eigenstates of Jones matrix on a Poincaré sphere (right panel) corresponding to the scenarios in (**c**). They correspond respectively to Solomon knot (**d**) in i subspace, Trefoil link (**h**) in ii-iii subspace, and Hopf link (**g**) in iv subspace, two EPs (**e**), and a single EP (**f**). Middle panels in (**d**), (**g**), (**h**) corresponding to braid diagrams with braid closured by the up and down boundaries.

Based on the theoretical underpinnings previously outlined, it becomes conceptually straightforward to design non-Hermitian metasurfaces with varied braiding topologies. The device is actualized using an array of terahertz plasmonic SRRs configured for coupled *y*- and *x*-polarized modes. The devised plasmonic non-Hermitian metasurface, detailed in Fig. 2a, comprises a dual-layer structure with semiconductor islands incorporated into the selected SRR gaps (see design details in the Supplementary Text). Injecting pump light into the semiconductor islands (200-nm-thick germanium) with 1.55 eV photon energy serves as a precise tuning knob to introduce non-Hermiticity through transient photocarrier-induced conductivity. Such conductivity short-circuits the capacitive gap and leads to non-radiative loss in the corresponding SRRs. Under varying pump fluence to facilitate controllable loss imbalance in the resonators, we mainly focus on the evolution of the knot invariant (knot invariant) and eigenvalue spectra. In experiments, we mainly monitor the responses from 0.45 to 0.85 THz, where the metasurface is designed to exhibit strong resonances. In this frequency window, experimentally accessible eigen-spectrum bounds are consistently determined with $\text{Im}(\lambda_{1,2}) = 0$. Consequently, while our experimental knot invariants manifest as one unit lower than their theoretical counterparts, they still effectively capture topological phase transitions by focusing on the change in knot invariant between pumped and non-pumped scenarios (refer to Supplementary Text for comprehensive details).

For more detailed manifestation of braiding topology variation, we particularly extract the eigen-spectra in the upper panels in Figs. 2b-d. The spectra capture braiding topology transitions, "Solomon link→Hopf link" for META 1, and "Solomon link→Trefoil knot" META 2 and META 3, which correspond to the subspace transitions from i to ii and i to iii in Fig. 1c, respectively. Figs. 2b-d demonstrate how integrating semiconductor islands into different gaps significantly diversifies braiding topology transitions. When the semiconductor islands are embedded into both 'p' and 'm' resonators, we observe the knot invariant variation ($\Delta \mathcal{W} = \mathcal{W}_{\text{pump}} - \mathcal{W}_{\text{no pump}}$) from 0 to -2 as a function of pump fluence, alongside crossing dual EP singularities (Fig. 2b). Similarly, employing a single control in either resonator 'p' or 'm' results in a knot invariant variation from 0 to -1, as shown in Figs. 2c-d. Corresponding EPs are noted at specific pump fluence levels where knot invariant alterations occur. Despite probing resonant modes in the far-field, our methodology reliably captures key braiding features across the resonant frequency range, providing a comprehensive portrayal of the braiding topology transitions in non-Hermitian metasurfaces.

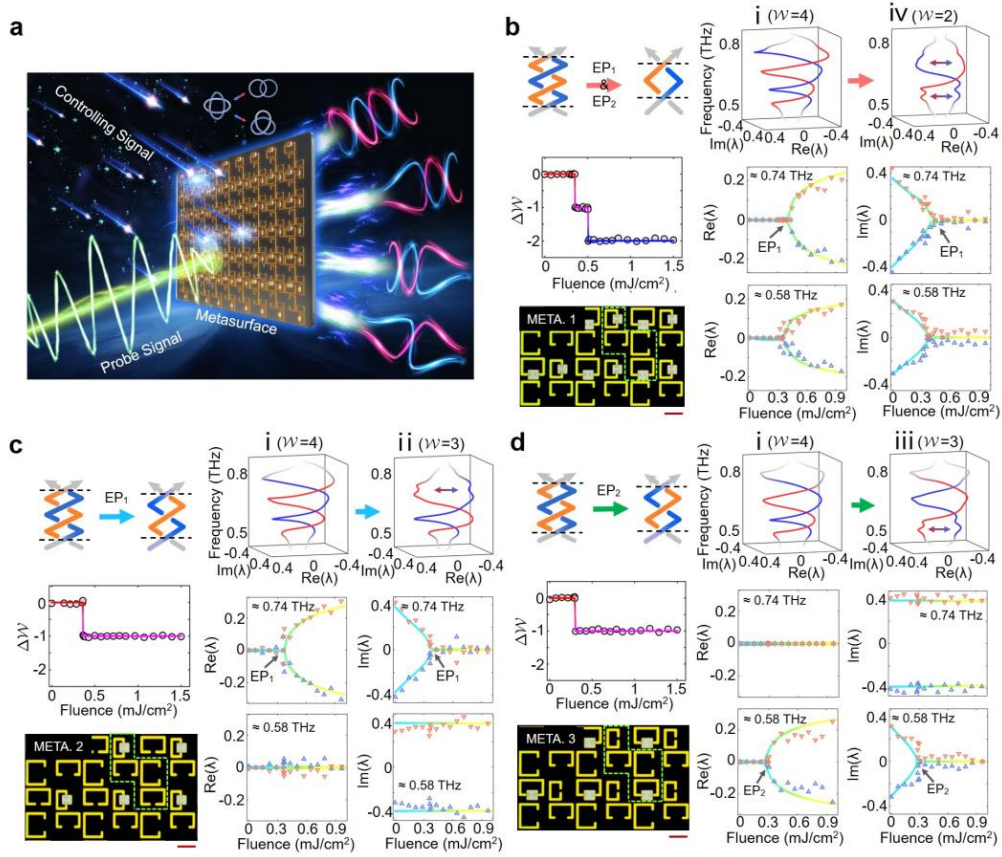

**Figure 2 | Braiding topology transitions in optically-reconfigurable non-Hermitian metasurfaces**. **a**, Schematic of the optically-reconfigurable metasurface, integrated with semiconductor micro-islands. Pump fluence of external signal light modulates the loss imbalance $\Delta\gamma'_{qn}$ and $\Delta\gamma'_{pm}$ between orthogonal resonances by adjusting the photocarriers in semiconductor islands, enabling diverse braiding topology. **b-d**, Experimental observation of braiding topology transition. Upper panel: eigenspectral responses representing braiding topology variation when pump with femtosecond laser 1.5mJ/cm$^2$. Middle left panel: modulation of braid invariants as a function of pump fluence, corresponding to the framework in Fig. 1c, varying from subspace i to ii, iii, and iv, respectively. Bottom left panel: optical microscopy images of the fabricated metasurfaces and single unit-cells are framed by green dotted lines, marked as META 1 (**b**), META 2 (**c**), and META 3 (**d**). Each unit cell comprising four distinct resonators, aligned vertically in descending order designated as 'p', 'm', 'q', and 'n', which respectively corresponds the resonators in Fig. 1b. A scale bar of 40 μm is included for size reference. Bottom right panels: extracted real and imaginary eigenvalues at two frequencies where EP$_1$ and EP$_2$ occur, respectively. Dots denote experimental data and curves are fitted for better visualization.

Further, we demonstrate the ability of the proposed metasurface to control the braiding topology at picosecond timescale. In our experiments, femtosecond pulses are divided into two parts of beams to act as a controlling signal and terahertz pulse generation, respectively, and their time delay is realized through displacement actuation (Fig. 3a). The amorphous germanium-hybrid metasurface, leveraging the limited lifetime of photocarriers (Fig. 3b), enables nearly instantaneous switching within picoseconds to femtoseconds [40]. When the semiconductor islands are inserted into the gap of an SRR, the non-radiative loss the SRR is then introduced by the photoexcitation since the photoconductivity can short-circuit the inductance-capacitance resonance to some extent. This attribute may facilitate the realization of an ultrafast non-Hermitian phase transition, leading to braiding topology transition. In our dynamic braiding topology characterization, an infrared femtosecond laser pulse activates the metasurface, and polarization-resolved terahertz pulses are used for probing. Various pump-probe time delays capture the full evolution of terahertz spectrum responses. Figs. 3c, f present the time-resolved knot invariant changes for three metasurfaces, showing complete On-Off-On switching cycles within 15 ps. Both the rise and fall times of the phase transition are shorter than the measuring limit in our system, highlighting the unstable and ultrasensitive nature of the exact EPs. Eigenvalue spectra, depicted in Figs. 3d, g, illustrate these transitions at key time moments. Notably, the post-switch-off recovered braiding spectra closely resemble the initial states, reaffirming the stability and rapidity of phase transitions. Further, in Figs. 3e, h, we analyze the complex eigenvalues at two EP frequencies, observing rapid alternation and recovery of level repulsion and degeneracy on picoseconds time scale. The EP singularities, marking the braiding topology phase transitions, are distinctly visible at knot invariant jump moments.

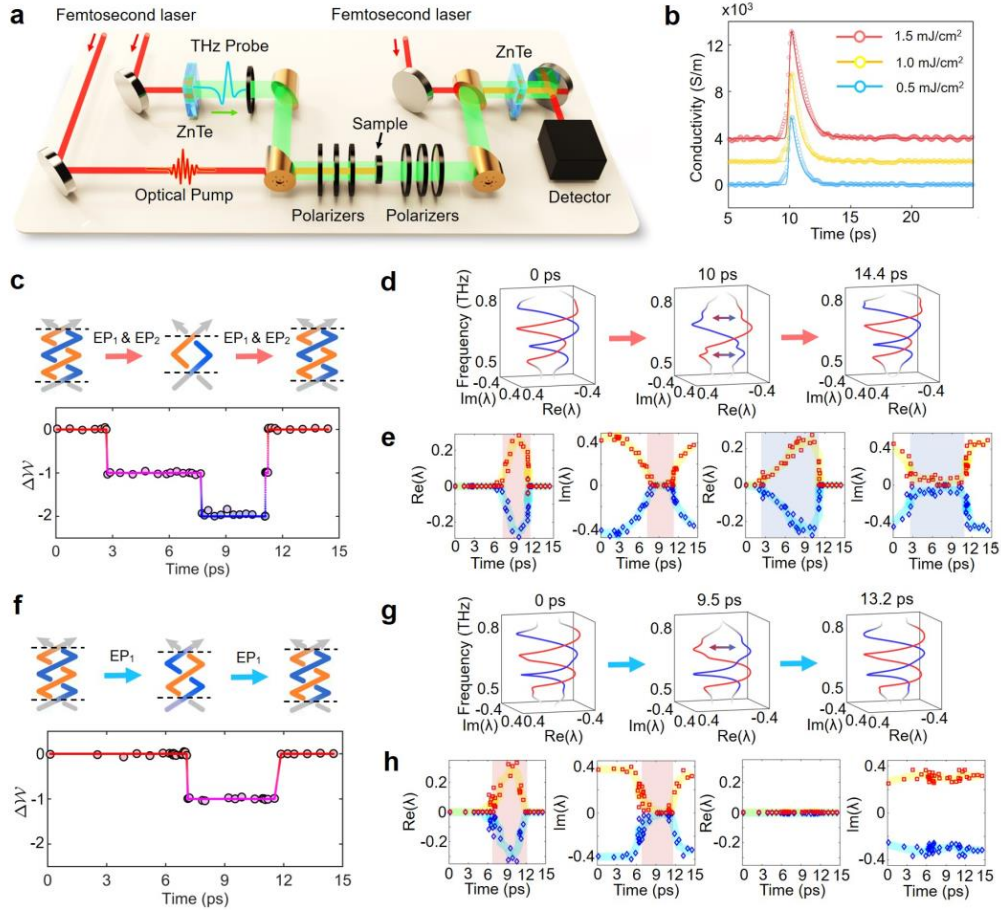

**Figure 3 | Observation of ultrafast switching of braiding topology. a**, Experimental setup consists of optical pump pulse and terahertz probe pulse with their time delay controlled by a variable displacement actuation. The terahertz pulse is focused on the sample which is excited by the optical pump and then the transmitted terahertz wave is recorded by another femtosecond laser beam according to electro-optical sampling method. **b**, Time-resolved photocarrier recombination dynamics within a 200-nm-thick layer of pure amorphous germanium film. This is deduced by the negative differential transmission ($-\Delta E/E_0$) of the terahertz wave, attributable to photoinduced conductivity of the semiconductor layer. The data demonstrates a complete recovery time of less than 5 picoseconds, highlighting the rapid response characteristics of the material. **c**, **f**, Braiding topology dynamics of the metasurfaces, pumped by an external signal light with a fluence of 1.5 mJ/cm². The terahertz pulse response is probed across various pump-probe time delays. **d**, **g**, Spectral eigenvalue responses at selected time intervals, illustrating the ultrafast processes of knot disentangling and reformation. **e**, **h**, Complete evolution of the real and imaginary parts of the complex eigenvalues, highlighting the EPs and transition processes within an ultrafast timeframe. Notably, each EP instance corresponds with the transition point.

Beyond topology of eigenvalues, braiding transitions involve concurrent topological changes in the associated eigenstates. Importantly, at close proximity to EPs, these eigenstate contours experience merging and reconfiguring, which will be experimentally confirmed. Depicting the system's eigenstates on the Poincaré sphere (Fig. 4) facilitates monitoring their evolution through transitions between different braids via the EPs. At spectral singularity, the eigenstates and eigenvalues coalesce, marked at the sphere's poles for left- or right-chiral eigenstates, respectively (Fig. 4, red and blue stars). As illustrated in the left two panels of Figs. 4 a-c, the eigenstates initially near the equator indicate linear polarization states. Upon photoexcitation, modes at frequencies of $f_{EP1}$ and $f_{EP2}$ close and converge towards the poles. At the chiral EPs, as depicted in the middle two panels of Figs. 4, eigenmodes from both branches merge into a singular mode. This coalescence and subsequent cross-connection around the chiral EPs signify a knot invariant shift [41]. Increasing pump fluence then diverges the modes back towards the equator predominantly in linear polarization states. This separation of eigenstates signifies a departure from the EPs after a reduction in the knot invariant. Incorporating the theoretical findings depicted in Fig. 1, we observe a striking concordance with the predicted phenomena in the eigenstate trajectories on the Poincaré sphere when traversing EP. A single loop structure evolves into a double-loop configuration upon crossing a single EP, yet reverts back to a single loop after passing through two EPs. Therefore, during the process of braiding topology transitions, the manipulation of polarization eigenstates emerges as a pivotal aspect, presenting a novel avenue for polarization control and potentially unlocking new applications in the realm of controllable polarization dynamics.

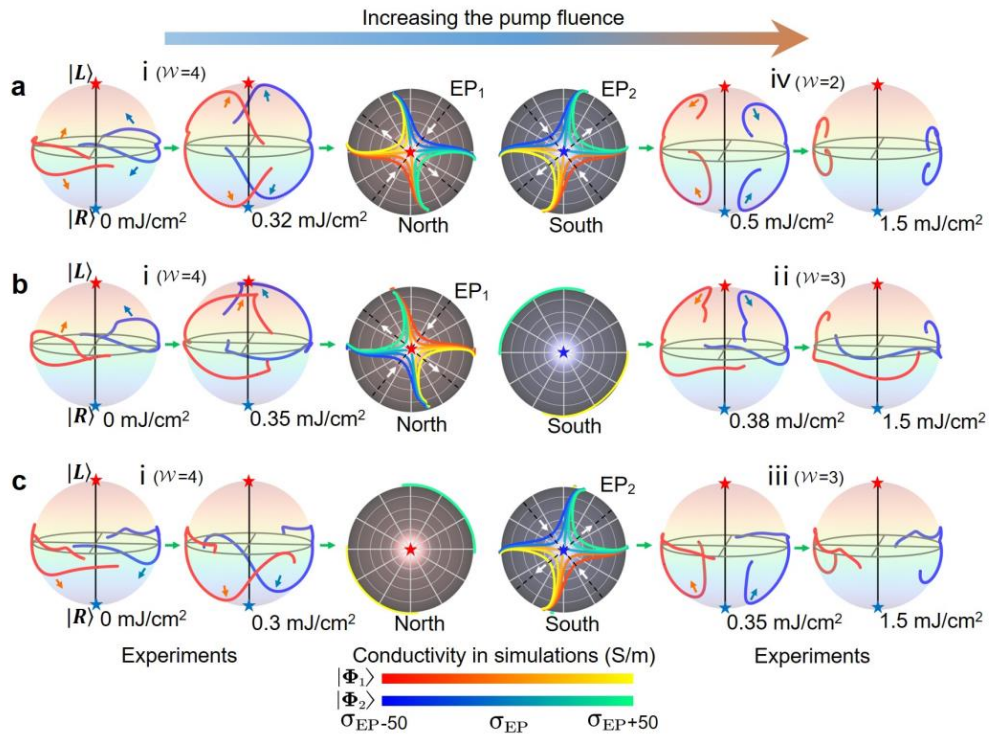

**Figure 4 | Evolution of the eigenpolarization states. a-c**, Eigenstates for META 1, META 2, and META 3. respectively, parametrically mapped onto the Poincaré sphere as a function of optical pump fluence. North and south poles are left $|L\rangle$ and right $|R\rangle$ circularly polarized eigenstates, respectively. Red and blue curves denote the eigenstates $|\Phi_1\rangle$ and $|\Phi_2\rangle$, respectively. Figure panels show: Experimental eigenstate mappings before encountering any EPs in the leftmost two panels; Numerical simulations illustrating the eigenstates in the vicinity of EPs, depicted in the middle set of panels; Experimental results of eigenstate configurations after traversing the singularity points, portrayed in the rightmost two panels. $\sigma_{EP}$ is the conductivity of semiconductor islands when EP occurs for the corresponding metasurface.

## Discussion

Our work has thus successfully demonstrated the ultrafast control of eigen-spectrum braiding topology transitions of the transmitted Jones matrices in non-Hermitian metasurfaces. Leveraging transient nonradiative loss modulation in semiconductor-hybrid terahertz plasmonic SRRs, we have experimentally evidenced various ultrafast transitions among the Solomon link, Trefoil knot, and Hopf link topologies. Different from conventional approaches where the braiding topology is defined in the momentum-space energy bandstructures or real-space field distributions, our approach introduces knot-theory principles to the scattering matrix formalism that governs the functionality of numerous optical devices. Our methodology thus fundamentally constitutes a critical tool for unraveling the intricate and non-trivial braiding configurations inherent in eigen-spectra. Looking ahead, the ultrafast control of the eigen-spectrum braids of the Jones matrices as well as the corresponding eigenpolarization states paves the way to topological optoelectronic devices for applications in advanced communications, sensing, and signal processing.

# Acknowledgments

Key Research and Development Program of the Ministry of Science and Technology grant 2022YFA1405200 (Y.Y.)

Key Research and Development Program of the Ministry of Science and Technology grant 2022YFA1404900 (Y.Y.)

Key Research and Development Program of the Ministry of Science and Technology grant 2022YFA1404704 (H.C.)

Key Research and Development Program of the Ministry of Science and Technology grant 2022YFA1404902 (H.C.)

National Natural Science Foundation of China (NNSFC) grant 62175215 (Y.Y.)

National Natural Science Foundation of China (NNSFC) grant 61975176 (H.C.)

National Natural Science Foundation of China (NNSFC) grant 62305384 (Y.H.)

National Natural Science Foundation of China (NNSFC) grant 62305298 (M.T.)

China National Postdoctoral Program for Innovative Talents grant BX20230310 (M.T.)

Youth Innovation Talent Incubation Foundation of National University of Defense Technology grant 2023-lxy-fhij-007 (Y.H.)

Key Research and Development Program of Zhejiang Province grant 2022C01036 (H.C.)

Fundamental Research Funds for the Central Universities grant 2021FZZX001-19 (Y.Y.)

Excellent Young Scientists Fund Program (Overseas) of China (Y.Y.)

# Author contributions

Y.Y., T.J., Y.H., and M.T. created the design. Y.H., M.T. and Y.Y. designed the experiment. Y.H. and M.T. fabricated samples. Y.H. carried out the measurement with the assistance from Z.R. and F.C. Y.H. analyzed the data. M.T. performed simulations, T.J., Y.Y., and H.C. provided the theoretical explanations. Y.H., M.T., Q.C. wrote the manuscript with the input from Y.Y., T.J., and H.C. Y.Y. supervised the project. All authors contributed extensively to this work.

# Competing interests

The authors declare no competing interests.

Data and materials availability

The data that support the findings of this study are available from the corresponding author upon reasonable request.

# References

1. Adams, C. C. The knot book (American Mathematical Society, Providence, RI, 2004), an elementary introduction to the mathematical theory of knots. *Revised reprint of the 1994 original*.
2. Thomson, W. II. On vortex atoms. The London, Edinburgh, and Dublin Philosophical Magazine and Journal of Science **34**, 15-24 (1867).
3. Hu, H. & Zhao, E. Knots and Non-Hermitian Bloch Bands. *Physical Review Letters* **126**, 010401, doi:10.1103/PhysRevLett.126.010401 (2021).
4. Witten, E. Quantum field theory and the Jones polynomial. *Communications in Mathematical Physics* **121**, 351-399, doi:10.1007/BF01217730 (1989).
5. Zheng, F. *et al.* Hopfion rings in a cubic chiral magnet. *Nature* **623**, 718-723, doi:10.1038/s41586-023-06658-5 (2023).
6. Wu, Q., Soluyanov, A. A. & Bzdušek, T. Non-Abelian band topology in noninteracting metals. *Science* **365**, 1273-1277, doi:10.1126/science.aau8740 (2019).
7. Cui, X. *et al.* Experimental realization of stable exceptional chains protected by non-Hermitian latent symmetries unique to mechanical systems. *arXiv preprint arXiv:2304.10347* (2023).
8. Tang, W. *et al.* Exceptional nexus with a hybrid topological invariant. *Science* **370**, 1077-1080, doi:10.1126/science.abd8872 (2020).
9. Leach, J., Dennis, M. R., Courtial, J. & Padgett, M. J. Knotted threads of darkness. *Nature* **432**, 165-165, doi:10.1038/432165a (2004).
10. Kong, L.-J. *et al.* High capacity topological coding based on nested vortex knots and links. *Nature Communications* **13**, 2705, doi:10.1038/s41467-022-30381-w (2022).
11. Zhang, X.-L. *et al.* Non-Abelian braiding on photonic chips. *Nature Photonics* **16**, 390-395, doi:10.1038/s41566-022-00976-2 (2022).
12. Noh, J. *et al.* Braiding photonic topological zero modes. *Nature Physics* **16**, 989-993, doi:10.1038/s41567-020-1007-5 (2020).
13. Iadecola, T., Schuster, T. & Chamon, C. Non-Abelian Braiding of Light. *Physical Review Letters* **117**, 073901, doi:10.1103/PhysRevLett.117.073901 (2016).
14. Wang, K. *et al.* Generating arbitrary topological windings of a non-Hermitian band. *Science* **371**, 1240-1245, doi:10.1126/science.abf6568 (2021).
15. Wang, K., Dutt, A., Wojcik, C. C. & Fan, S. Topological complex-energy braiding of non-Hermitian bands. *Nature* **598**, 59-64, doi:10.1038/s41586-021-03848-x (2021).

16   Wang, D., Wu, Y., Zhang, Z. Q. & Chan, C. T. Non-Abelian Frame Charge Flow in Photonic Media. *Physical Review X* **13**, 021024, doi:10.1103/PhysRevX.13.021024 (2023).

17   Yang, E. *et al.* Observation of Non-Abelian Nodal Links in Photonics. *Physical Review Letters* **125**, 033901, doi:10.1103/PhysRevLett.125.033901 (2020).

18   Dennis, M. R., King, R. P., Jack, B., O'Holleran, K. & Padgett, M. J. Isolated optical vortex knots. *Nature Physics* **6**, 118-121, doi:10.1038/nphys1504 (2010).

19   Kleckner, D. & Irvine, W. T. M. Creation and dynamics of knotted vortices. *Nature Physics* **9**, 253-258, doi:10.1038/nphys2560 (2013).

20   Larocque, H. *et al.* Reconstructing the topology of optical polarization knots. *Nature Physics* **14**, 1079-1082, doi:10.1038/s41567-018-0229-2 (2018).

21   Pisanty, E. *et al.* Knotting fractional-order knots with the polarization state of light. *Nature Photonics* **13**, 569-574, doi:10.1038/s41566-019-0450-2 (2019).

22   Wang, H. & Fan, S. Photonic Spin Hopfions and Monopole Loops. *Physical Review Letters* **131**, 263801, doi:10.1103/PhysRevLett.131.263801 (2023).

23   Larocque, H. *et al.* Optical framed knots as information carriers. *Nature Communications* **11**, 5119, doi:10.1038/s41467-020-18792-z (2020).

24   Dai, T. *et al.* Non-Hermitian topological phase transitions controlled by nonlinearity. *Nature Physics*, doi:10.1038/s41567-023-02244-8 (2023).

25   Raja, A. S. *et al.* Ultrafast optical circuit switching for data centers using integrated soliton microcombs. *Nature Communications* **12**, 5867, doi:10.1038/s41467-021-25841-8 (2021).

26   Rutckaia, V. & Schilling, J. Ultrafast low-energy all-optical switching. *Nature Photonics* **14**, 4-6, doi:10.1038/s41566-019-0571-7 (2020).

27   Grinblat, G. *et al.* Efficient ultrafast all-optical modulation in a nonlinear crystalline gallium phosphide nanodisk at the anapole excitation. *Science Advances* **6**, eabb3123, doi:10.1126/sciadv.abb3123.

28   Hui, D. *et al.* Ultrafast optical switching and data encoding on synthesized light fields. *Science Advances* **9**, eadf1015, doi:10.1126/sciadv.adf1015.

29   Shaltout, A. M., Shalaev, V. M. & Brongersma, M. L. Spatiotemporal light control with active metasurfaces. *Science* **364**, eaat3100, doi:10.1126/science.aat3100 (2019).

30   Gu, J. *et al.* Active control of electromagnetically induced transparency analogue in terahertz metamaterials. *Nature Communications* **3**, 1151, doi:10.1038/ncomms2153 (2012).

31   Zhang, S. *et al.* Photoinduced handedness switching in terahertz chiral metamolecules. *Nature Communications* **3**, 942, doi:10.1038/ncomms1908 (2012).

32   Song, Q., Odeh, M., Zúñiga-Pérez, J., Kanté, B. & Genevet, P. Plasmonic topological metasurface by encircling an exceptional point. *Science* **373**, 1133-1137, doi:10.1126/science.abj3179 (2021).

33  Lawrence, M. *et al.* Manifestation of PT Symmetry Breaking in Polarization Space with Terahertz Metasurfaces. *Physical Review Letters* **113**, 093901, doi:10.1103/PhysRevLett.113.093901 (2014).

34  Baek, S. *et al.* Non-Hermitian chiral degeneracy of gated graphene metasurfaces. *Light: Science & Applications* **12**, 87, doi:10.1038/s41377-023-01121-6 (2023).

35  Hatano, N. & Nelson, D. R. Localization Transitions in Non-Hermitian Quantum Mechanics. *Physical Review Letters* **77**, 570-573, doi:10.1103/PhysRevLett.77.570 (1996).

36  Nakamura, D., Inaka, K., Okuma, N. & Sato, M. Universal Platform of Point-Gap Topological Phases from Topological Materials. *Physical Review Letters* **131**, 256602, doi:10.1103/PhysRevLett.131.256602 (2023).

37  Ding, K., Fang, C. & Ma, G. Non-Hermitian topology and exceptional-point geometries. *Nature Reviews Physics* **4**, 745-760, doi:10.1038/s42254-022-00516-5 (2022).

38  Shen, H., Zhen, B. & Fu, L. Topological Band Theory for Non-Hermitian Hamiltonians. *Physical Review Letters* **120**, 146402, doi:10.1103/PhysRevLett.120.146402 (2018).

39  Kauffman, L. H. *Knots and physics*. Vol. 1 (World scientific, 2001).

40  Lim, W. X. *et al.* Ultrafast All-Optical Switching of Germanium-Based Flexible Metaphotonic Devices. *Advanced Materials* **30**, 1705331, doi:https://doi.org/10.1002/adma.201705331 (2018).

41  Ergoktas, M. S. *et al.* Topological engineering of terahertz light using electrically tunable exceptional point singularities. *Science* **376**, 184-188, doi:10.1126/science.abn6528 (2022).